%% file: main.tex
\documentclass[sigconf,9pt]{acmart}

\acmYear{2026}\copyrightyear{2026}
\setcopyright{cc}
\setcctype[4.0]{by}
\acmConference[NAIC '26]{ACM SIGCOMM Workshop on Networks for AI Computing}{August 17--21, 2026}{Denver, CO, USA}
\acmBooktitle{ACM SIGCOMM Workshop on Networks for AI Computing (NAIC '26), August 17--21, 2026, Denver, CO, USA}
\acmDOI{10.1145/3789240.3828750}
\acmISBN{979-8-4007-2467-1/26/08}

\usepackage{algorithm}
\usepackage{algpseudocode}
\usepackage{bm}
\usepackage{booktabs}              
\usepackage{enumitem}
\usepackage{xspace}                
\usepackage{subcaption}
\usepackage{xcolor}
\usepackage{tabularx}
\usepackage{marginnote}
\AtBeginDocument{}

\DeclareFontShape{T1}{cmtt}{b}{n} {<->ssub * cmtt/m/n}{}
\DeclareFontShape{T1}{cmtt}{b}{it}{<->ssub * cmtt/m/it}{}
\DeclareFontShape{T1}{cmtt}{b}{sl}{<->ssub * cmtt/m/sl}{}
\DeclareFontShape{T1}{cmtt}{bx}{n} {<->ssub * cmtt/m/n}{}
\DeclareFontShape{T1}{cmtt}{bx}{it}{<->ssub * cmtt/m/it}{}
\DeclareFontShape{T1}{cmtt}{bx}{sl}{<->ssub * cmtt/m/sl}{}

\newcommand{\system}{ARK\xspace} 

\definecolor{darkgreen}{RGB}{0,128,0}
\newcommand{\ahmed}[1]{}

\title{\system: Avoiding Routing Collisions for KV Cache Transfer in Disaggregated LLM Inference}

\author{Hung-Chun Lin}
\authornote{Both authors contributed equally to this research.}
\affiliation{%
  \institution{Georgia Institute of Technology}
  \city{Atlanta}\state{GA}\country{USA}}
\email{hlin464@gatech.edu}

\author{Ting-Wei Hsu}
\authornotemark[1]
\affiliation{%
  \institution{Georgia Institute of Technology}
  \city{Atlanta}\state{GA}\country{USA}}
\email{thsu49@gatech.edu}

\author{Chung-En Ho}
\affiliation{%
  \institution{Georgia Institute of Technology}
  \city{Atlanta}\state{GA}\country{USA}}
\email{cho322@gatech.edu}

\author{Ahmed Saeed}
\affiliation{%
  \institution{Georgia Institute of Technology}
  \city{Atlanta}\state{GA}\country{USA}}
\email{asaeed@cc.gatech.edu}

\renewcommand{\shortauthors}{Lin et al.}

\begin{document}

\begin{abstract}

Disaggregated LLM inference separates the prefill and decode phases across GPU pools, generating massive KV-cache transfers. Because these transfers last hundreds of milliseconds to seconds, they behave as mega elephant flows that dominate link bandwidth utilization. In this regime, stateless ECMP can perform poorly: hash collisions may overload one spine link while leaving others idle, stretching transfer times by seconds. Yet this same persistence makes coordination practical. Since these flows are long-lived, even lightweight one-to-all coordination can be amortized over their lifetime. We present ARK, a distributed elephant-flow path reservation mechanism. ARK coordinates senders to choose source ports whose hashes map concurrent flows onto distinct spines, without requiring switch changes or receiver-side packet reordering. Packet-level RDMA simulations show that ARK reduces mean and P95 FCT by up to 27.3\% and 34.0\% under moderate load, and further reduces mean TTFT by up to 12.9\%.
\end{abstract}



\maketitle

\input{introduction}
\input{background}
\input{method}
\input{evaluation}
\input{conclusion}

\input{ethics}

\bibliographystyle{ACM-Reference-Format}
\bibliography{reference}

\end{document}

%% file: introduction.tex
\section{Introduction}

The rise of large language models (LLMs) is reshaping data-center utilization. Modern LLM serving pipelines split execution into two phases with disjoint resource profiles: a compute-bound \emph{prefill} phase that ingests the user prompt, and a memory-bandwidth-bound \emph{decode} phase that emits tokens autoregressively. Co-locating both phases on the same GPU pool over-provisions one resource for the other, motivating recent serving systems---DistServe~\cite{zhong2024distserve} and Splitwise~\cite{patel2024splitwise}---to physically \emph{disaggregate} prefill from decode across separate GPU pools. 
The price of this architecture is a new dominant communication pattern: the key--value (KV) cache produced by a prefill GPU must be shipped over the data-center fabric to the decode GPU before token generation can begin. Open-source inference platforms such as vLLM~\cite{kwon2023efficient}, SGLang~\cite{zheng2024sglang}, NVIDIA Dynamo~\cite{nvidia2025dynamo}, and llm-d~\cite{llmd2025} already implement this prefill--decode disaggregation pattern, and major LLM operators have adopted it at production scale: Moonshot AI's Kimi runs on the Mooncake disaggregated architecture~\cite{qin2025mooncake}, and DeepSeek-V3 is served via SGLang's disaggregated reference deployment~\cite{lmsys2025deepseek_ep}.

KV cache flows are unlike anything the fabric has seen. Where a traditional data-center ``elephant flow'' typically peaks around tens of megabytes~\cite{song2023conweave}, a single 100K-token KV cache for LLaMA-3.1-405B already reaches 48~GB.\footnote{KV-cache footprint per token is $2 \cdot L_h \cdot H_{kv} \cdot d_h \cdot s$, where $L_h$ is the number of decoder layers, $H_{kv}$ the number of KV heads, $d_h$ the head dimension, and $s$ the precision in bytes. For LLaMA-3.1-405B (126 layers, 8 KV heads, $d_h{=}128$, FP16), it has 504~KB per token.} A single such flow can saturate a 400~Gbps link for 960~ms. Because this transfer sits on the critical path of every request, even a small increase in flow completion time (FCT) directly amplifies time-to-first-token (TTFT), potentially leading to Service Level Objective (SLO) violations.

Production fabrics commonly use Equal-Cost Multi-Path (ECMP) routing to spread flows across redundant equal-cost paths.
ECMP selects a next hop by hashing the packet 5-tuple (source IP, destination IP, source port, destination port, and protocol), thereby pinning each flow to a single path.
Modern AI cluster fabrics retain ECMP rather than replace it: Meta's RoCE deployment for distributed training~\cite{gangidi2024rdma} and Alibaba HPN~\cite{qian2024hpn} both keep commodity ECMP and engineer around its weaknesses, and the production inference stacks surveyed in \S\ref{sec:background} ship on the same baseline.
The hash, however, is oblivious to flow size, so two KV cache mega-flows that happen to hash to the same spine collide and share its capacity for the multi-second lifetime of the transfer, 
potentially leaving other paths idle. Existing techniques for avoiding such collisions in the network~\cite{alfares2010hedera, alizadeh2014conga, vanini2017letflow, qureshi2022plb, song2023conweave} introduce significant cost either through hardware changes or specific software requirements.

We propose \system,
a lightweight distributed elephant-flow path reservation mechanism that runs entirely in commodity end-host software while leaving the switch data plane untouched. Each host maintains a replicated reservation table recording which spines are currently occupied by active KV cache transfers; when a new flow arrives, the local controller selects an unreserved spine and dispatches the flow with a precomputed source-port offset that maps to it under ECMP, and broadcasts the reservation to peers via a low-bandwidth control channel. Because the mechanism only manipulates the source port---a field every host already controls---the data plane requires no switch modification and no packet reordering at the receiver. We evaluate \system\ in a packet-level RDMA simulator (SimAI/NS-3) driven by realistic disaggregated-inference workloads from Vidur~\cite{agrawal2024vidur}. In a two-tier fat-tree of 16 GPUs (8 prefill / 8 decode) under a LLaMA-3-8B Zipf workload, \system\ reduces mean / P95 flow completion time (FCT) by up to 27.3\% / 34.0\% and end-to-end time-to-first-token (TTFT) by up to 12.9\%, with the largest gains in the load regime where the leaf--spine fabric is congested but not yet saturated.

%% file: background.tex
\begin{figure*}[!t]
    \centering
    \includegraphics[width=2.1\columnwidth]{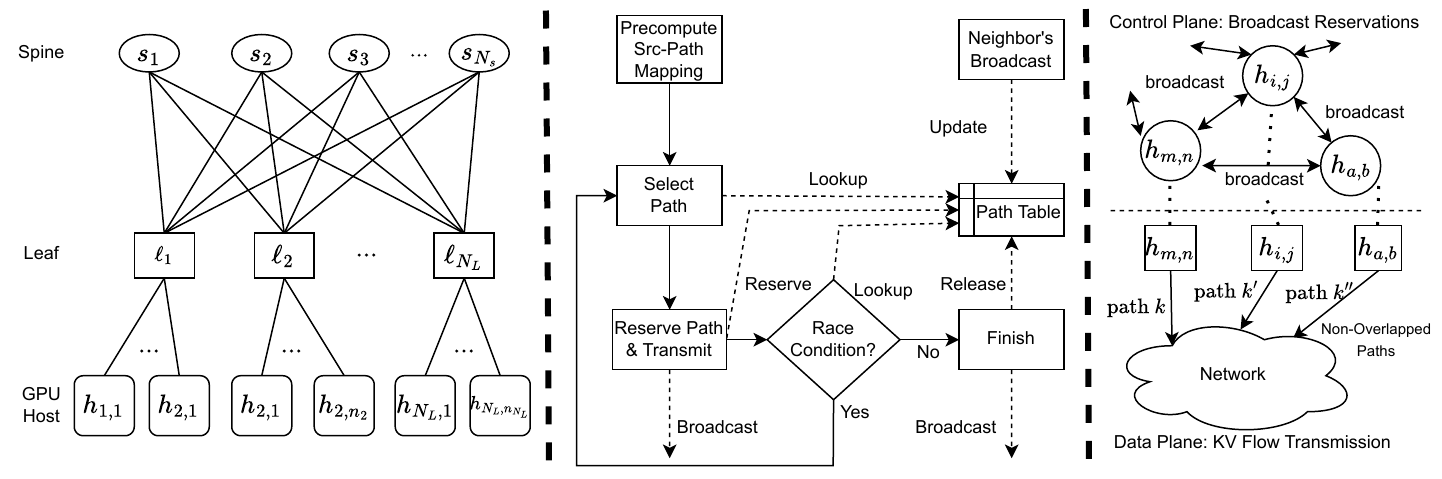}
    \caption{(a) Network topology (b) \system~path reservation controller (c) \system~deployment with separated control and data planes}
    \label{fig:graph_topology}
\end{figure*}

\section{Background and Motivation}
\label{sec:background}

\noindent{\bf Prefill--decode disaggregation.}
LLM serving alternates between compute-bound \emph{prefill} (attention $\Theta(L^2)$ in FLOPs) and high-bandwidth memory (HBM) bandwidth-bound \emph{decode} (each iteration reads the full KV cache from HBM)~\cite{patel2024splitwise,zhong2024distserve}. Co-locating both phases either inflates inter-token latency through prefill--decode interference or forces a coupled pool to over-provision against the independent TTFT and time-per-output-token (TPOT) service-level objectives (SLOs), each gated by a different resource axis~\cite{zhong2024distserve}. Disaggregating onto separate pools lets each pool be sized to its own bottleneck, optionally with a different accelerator model per phase~\cite{patel2024splitwise}.
Disaggregated serving is now common practice. Mooncake~\cite{qin2025mooncake}, which serves Moonshot AI’s Kimi chatbot, operates disaggregated KV-cache pipelines at production scale, while the SGLang reference deployment of DeepSeek-V3 on 96 H100 GPUs adopts prefill--decode disaggregation~\cite{lmsys2025deepseek_ep}. In the open-source ecosystem, both vLLM~\cite{kwon2023efficient} and SGLang~\cite{zheng2024sglang} support disaggregated serving. More recently, two industry-led efforts---NVIDIA Dynamo~\cite{nvidia2025dynamo} and llm-d~\cite{llmd2025}---have made prefill--decode disaggregation a central architectural principle.

\noindent{\bf Application-level KV cache optimizations}
Once prefill--decode disaggregation is taken as given, two orthogonal lines of work address the resulting communication cost: one above the fabric, on the KV cache itself, and one inside the fabric, on how its bytes are routed. \emph{Multi-tier KV stores} externalize the cache into pooled memory hierarchies so state can be shared across machines and reused across requests---MemServe~\cite{hu2024memserve} provides an elastic memory pool with locality-aware routing, and LMCache~\cite{cheng2025lmcache} exposes a cross-engine KV layer for transfer between different serving entities. A complementary \emph{transfer-optimization} thread shrinks or reroutes the bytes themselves: CacheGen~\cite{liu2024cachegen} compresses and adaptively streams the cache to shrink bytes-on-wire, DualPath~\cite{wu2026dualpath} routes reads through decode engines to rebalance storage NIC load, and Yue~et~al.~\cite{yue2025rtt} analyze per-block protocol overhead in geo-distributed deployments. None of these approaches controls \emph{which} equal-cost path the remaining bytes traverse once admitted to the fabric---a dimension that ARK addresses and that compounds with, rather than competes against, the bytes-on-wire reductions above. We turn next to how production fabrics balance the bytes that do hit the wire.

\noindent{\bf Load balancing in modern AI fabrics.}
Industry has demonstrated willingness to build custom hardware and software for AI workloads whenever the cost-benefit warrants it. For elephant-flow load balancing on AI fabrics, however, the deployed reality is uniformly commodity ECMP. Meta's RoCE deployment for distributed training~\cite{gangidi2024rdma} chose to enrich the input space (``Enhanced ECMP'') and layer centralized traffic engineering rather than replace ECMP outright; Alibaba HPN~\cite{qian2024hpn} kept ECMP and rebuilt the topology as a dual-plane, dual top-of-rack (ToR) fabric to suppress hash polarization. 

The relevant question for an in-network elephant-flow mechanism is therefore comparative cost on top of this baseline, not technical feasibility. CONGA~\cite{alizadeh2014conga} adds custom congestion-aware leaf ASICs across the fabric; ConWeave~\cite{song2023conweave} requires programmable ToRs paired with per-flow reordering buffers; LetFlow~\cite{vanini2017letflow} adds switch-level flowlet detection. While packet spraying effectively mitigates ECMP collisions, it delivers packets out of order; for mega-flows like KV cache transfers, reassembling them imposes a heavy CPU burden at the receiver. Host-side PLB~\cite{qureshi2022plb} avoids switch upgrades but is \emph{reactive}: it only repaths after $M$ consecutive ECN-marked rounds, and only when the connection is idle, since mid-flow IPv6 Flow Label changes cause reordering that RoCEv2 RNICs misread as loss~\cite{song2023conweave}. This reactive, congestion-control-first posture matches TCP workloads where flow size and arrival are unknown, but it is a poor fit for KV cache transfers, whose size and destination are known the instant the prefill engine finalizes the cache. Worse, RDMA's hardware-paced KV transfers leave essentially no idle gaps~\cite{song2023conweave}, so PLB's safe-repath precondition rarely fires within a KV flow's lifetime. \system instead exploits this admission-time knowledge (\S\ref{subsec:integration}) to fix the path \emph{before the first byte ships}, sidestepping congestion-control feedback, reactive lag, and the idle-gap requirement.

\noindent{\bf Opportunity for coordination.}
\label{subsec:gap}
KV cache flows differ from the flows that motivated prior datacenter load balancers along two axes that, together, expose a design space earlier mechanisms could not exploit.

\noindent \textit{Flow lifetime is long enough to amortize coordination.} A 1.6~GB mean KV transfer occupies a 100~Gbps spine link for at least 128~ms, and longer when sharing the spine---three to four orders of magnitude above the millisecond-class flows targeted by reactive flowlet schemes~\cite{vanini2017letflow,qureshi2022plb}. Millisecond-scale admission-time coordination therefore amortizes well under 1\% of flow lifetime, where reactive flowlet schemes have no comparable budget.

\noindent \textit{Each flow's identity is known at admission.} The destination GPU, byte count, and start time of every KV transfer are determined the moment the prefill engine finalizes the cache; no runtime elephant detection (as in Hedera~\cite{alfares2010hedera}) is required. The equal-cost paths are likewise enumerable at the host, so both the flow and its paths are exactly known before any byte ships.

\system\ exploits both. At admission, a local controller picks an unreserved spine from a host-replicated reservation table, dispatches the flow with a precomputed source-port offset that hashes onto it, and broadcasts the reservation to peers. The decision needs no runtime elephant detection and no in-flight congestion signaling (CONGA~\cite{alizadeh2014conga}); each flow uses a single path, so no receiver-side reordering is required.

%% file: method.tex
\section{ARK System Design}
\label{sec:approach}

\system~enables ECMP hash collision avoidance through a distributed path reservation controller on each server.
By maintaining a local replica of the reservation table via lightweight broadcasts between peers, each controller is aware of which paths are currently available when it admits an outgoing KV cache flow.

\noindent{\bf \system\ as minimal scheduling.}
Avoiding ECMP collisions is fundamentally a scheduling problem, and \system\
sits at the lightweight end of a wide design space. Centralized schedulers
coordinate the whole fabric through an arbiter---Fastpass~\cite{perry2014fastpass}
per packet, Flowtune~\cite{perry2017flowtune} per flowlet---while a more recent
line schedules \emph{collective} communication for AI training
(FAST~\cite{lei2026fast} for MoE \texttt{alltoallv}, Crux~\cite{cao2024crux} for
training collectives). \system\ occupies the opposite extreme by design: it
needs no central arbiter and no traffic matrix, scheduling only the choice of
spine, once per flow at admission. This is a fit to the long-lived,
admission-known KV workload, not a shortcut.

\noindent \textbf{Assumptions.}
We make two assumptions. First, the KV cache transfer application uses a single, well-known destination port across all peers, leading to fixed src IP, dst IP, dst port, and IP protocol per peer pair, so only the source port varies (the precompute below relies on this; multi-dport scenarios extend trivially to a per-(peer, dport) table). Second, host clocks are synchronized to within the broadcast delay (a few microseconds in our deployment), achievable on commodity datacenter fabrics via PTP~\cite{ieee1588}. This assumption is needed by the race-condition resolution rule in \S\ref{subsec:integration}.

\subsection{Path Reservation Controller}
\label{subsec:path_reservation_controller}

\noindent \textbf{Topology:}
We model the disaggregated LLM inference fabric as a two-tier spine-leaf network~\cite{fattree} (Fig.~\ref{fig:graph_topology}(a)) consisting of $N_S$ spine switches $V_S = \{s_1, \ldots, s_{N_S}\}$, $N_L$ leaf (top-of-rack) switches $V_L = \{\ell_1, \ldots, \ell_{N_L}\}$, and a set of GPU hosts $V_H$ in which $h_{i,m}$ denotes the $m$-th host attached to leaf $\ell_i$. Each leaf is fully connected to every spine, and every host has a single uplink to its leaf. For any cross-leaf host pair $(h_{i,m}, h_{j,n})$, the fabric exposes $K = N_S$ equal-cost paths
\begin{equation}
\begin{aligned}
    \mathcal{P}(h_{i,m}, h_{j,n}) \;&=\; \{p_1, \ldots, p_K\}, \\
    p_k \;&=\; (h_{i,m} \to \ell_i \to s_k \to \ell_j \to h_{j,n}),
\end{aligned}
    \label{eq:paths}
\end{equation}
each traversing exactly one spine. For derivation convenience we present the formalism on a two-tier spine-leaf instance, taking Alibaba HPN~\cite{qian2024hpn} as a representative production example, while \system\ extends to deeper Clos fabrics such as Meta's multi-tier RoCE deployment~\cite{gangidi2024rdma} by inverting the ECMP hash at each ECMP decision along the path. 

\noindent \textbf{ECMP routing and hash collisions:}
Switches forward each flow $f_i$ using standard ECMP, hashing the 5-tuple $\mathbf{q}_i = (\mathrm{srcIP}_i,\allowbreak \mathrm{dstIP}_i,\allowbreak \mathrm{srcPort}_i,\allowbreak \mathrm{dstPort}_i,\allowbreak \mathrm{proto}_i)$ and selecting
\begin{equation}
    \pi_i \;=\; p_{k_i}, \qquad k_i \;=\; \mathcal{H}(\mathbf{q}_i) \bmod K,
    \label{eq:ecmp_hash}
\end{equation}
where $\mathcal{H}$ is the switch hash function. Because $\mathcal{H}$ is oblivious to flow size, two elephant flows whose hashes coincide modulo $K$ share the same spine link and split its bandwidth, regardless of whether sibling spines are idle. AI Clos fabrics commonly oversubscribe the leaf--spine layer for cost reasons, which further amplifies collision impact: with less spare bandwidth, collision-induced queueing fills buffers faster. Their flow completion times $T_i^{\mathrm{FCT}}$ scale roughly with the number of competing flows on that link, with RDMA retransmissions further amplifying the penalty when buffers overflow. We write $\tau_i$ for the time at which flow $f_i$ is injected into the fabric.

\noindent \textbf{Reservation table:}
Each host runs a path reservation controller that maintains a local
path reservation table $\mathcal{R}(t) \subseteq \mathcal{P}$,
recording the paths currently occupied by active elephant flows across
the fabric.

\noindent \textbf{Boot-time precompute:}
Under this assumption, the controller precomputes a per-peer src
port-to-path mapping
\begin{equation}
    \mathcal{M}_{h_j}(p_k) \;=\;
        \bigl\{\, \delta \;:\;
            \mathcal{H}(\mathbf{q}^{\delta}) \bmod K = k \,\bigr\},
    \label{eq:precompute}
\end{equation}
once at server boot, where $p_k \in \mathcal{P}(\mathrm{self}, h_j)$ is
the $k$-th path toward $h_j$ and $\mathcal{M}_{h_j}(p_k)$ enumerates
the src-port offsets that route flows toward $h_j$ onto $p_k$. 

\noindent \textbf{Flow arrival:}
When a new flow $f_i$ arrives at the host at time $\tau_i$, the
controller selects an unreserved path and reads any one offset from
the corresponding bucket:
\begin{equation}
    \pi_i^{\mathrm{ARK}} \;=\; p_{k_i^*},
    \quad
    k_i^* \;=\; \min\bigl\{\, k \;:\; p_k \notin \mathcal{R}(\tau_i) \,\bigr\},
    \label{eq:ark_select}
\end{equation}
a short scan of $\mathcal{R}$ that probes at most $K$ paths. The corresponding source-port offset $\delta_i^{*}$ is then taken from the precomputed bucket $\mathcal{M}_{\mathrm{dst}_i}(p_{k_i^*})$. Upon assignment, the chosen path is added to the local path reservation table and broadcast to all peers:
\begin{equation}
    \mathcal{R}(\tau_i^+) \;=\; \mathcal{R}(\tau_i) \cup
        \bigl\{\, p_{k_i^*} \,\bigr\}.
\end{equation}
When the flow completes, the path is released from the table and a
corresponding RELEASE message is broadcast:
\begin{equation}
    \mathcal{R}\bigl((\tau_i + T_i^{\mathrm{FCT}})^+\bigr)
        \;=\; \mathcal{R}\bigl(\tau_i + T_i^{\mathrm{FCT}}\bigr)
              \setminus \bigl\{\, p_{k_i^*} \,\bigr\}.
\end{equation}
Peer controllers apply incoming RESERVE/RELEASE updates to their own path reservation tables, keeping them eventually consistent across the fabric.
When all paths are reserved, i.e., $|\mathcal{R}(t)| = K$, the controller falls back to vanilla ECMP by randomly drawing a path index $k_i^{*} \in \{1, \ldots, K\}$ and taking $\delta_i^{*}$ from $\mathcal{M}_{\mathrm{dst}_i}(p_{k_i^{*}})$.
The full procedure is summarized in Algorithm~\ref{alg:ark}.

\begin{algorithm}[t]
\footnotesize
\caption{\system~Path Reservation Controller}
\label{alg:ark}
\begin{algorithmic}[1]
\Require Path set $\mathcal{P}$; switch hash $\mathcal{H}(\cdot)$;
         host identifier $\mathrm{id}$; peer set $\mathcal{V}$
\State $\mathcal{R} \gets \emptyset$
        \Comment{$p_k \mapsto (t,\,\mathrm{id})$}
\Statex
\Procedure{Initialize}{} \Comment{run once at server boot}
    \For{each peer $h_j \in \mathcal{V}$}
        \State $\mathcal{M}_{h_j}[p_k] \gets \emptyset$ for all $p_k \in \mathcal{P}(\mathrm{self}, h_j)$
        \For{$\delta = 0, 1, \ldots, \delta_{\max}$}
            \State $\mathbf{q}^{\delta} \gets$ 5-tuple to $h_j$ with src port $= \delta$
            \State $k \gets \mathcal{H}(\mathbf{q}^{\delta}) \bmod K$
            \State $\mathcal{M}_{h_j}[p_k] \gets \mathcal{M}_{h_j}[p_k] \cup \{\delta\}$
        \EndFor
    \EndFor
\EndProcedure
\Statex
\Procedure{OnFlowArrival}{$f_i$}
    \State $t_i \gets$ current timestamp
    \If{$|\mathcal{R}| = K$}
        \Comment{Fall back to ECMP}
        \State $k_i^{*} \gets$ random $k \in \{1, \ldots, K\}$;\;
               $\delta_i^{*} \gets$ any $\delta \in \mathcal{M}_{\mathrm{dst}_i}[p_{k_i^{*}}]$
    \Else
        \State $k_i^{*} \gets \min\{k : p_k \notin \mathcal{R}\}$
        \State $\delta_i^{*} \gets$ any $\delta \in \mathcal{M}_{\mathrm{dst}_i}[p_{k_i^{*}}]$
        \State $\mathcal{R}[p_{k_i^{*}}] \gets (t_i,\,\mathrm{id})$
        \State \textbf{Broadcast}
               $(p_{k_i^{*}},\,\text{RESERVE},\,t_i,\,\mathrm{id})$ to peers
    \EndIf
    \State $\pi_i \gets p_{k_i^{*}}$
    \State \textbf{Create} RDMA queue pair for $f_i$ with src port $\delta_i^{*}$
    \State \textbf{Register} \Call{OnFlowCompletion}{$f_i, k_i^{*}$} callback
\EndProcedure
\Statex
\Procedure{OnFlowCompletion}{$f_i, k_i^{*}$}
    \State $\mathcal{R} \gets \mathcal{R} \setminus \{p_{k_i^{*}}\}$
    \State \textbf{Broadcast} $(p_{k_i^{*}},\,\text{RELEASE})$ to peers
\EndProcedure
\Statex
\Procedure{OnRemoteUpdate}{$p_k,\,\text{op},\,t_p,\,\mathrm{id}_p$}
    \If{$\text{op} = \text{RELEASE}$}
        \State $\mathcal{R} \gets \mathcal{R} \setminus \{p_k\}$
    \ElsIf{$p_k \notin \mathcal{R}$}
        \State $\mathcal{R}[p_k] \gets (t_p,\,\mathrm{id}_p)$
    \ElsIf{$(t_p,\,\mathrm{id}_p) < \mathcal{R}[p_k]$}
        \Comment{Race condition}
        \State $\mathcal{R}[p_k] \gets (t_p,\,\mathrm{id}_p)$
        \State \Call{OnFlowArrival}{$f$}
            \Comment{Re-hash local flow on $p_k$}
    \EndIf
\EndProcedure
\end{algorithmic}
\end{algorithm}

\subsection{ARK Deployment}
\label{subsec:integration}

Each host runs the controller of \S\ref{subsec:path_reservation_controller} as user-space code. Following Fig.~\ref{fig:graph_topology}(c), \system~separates the two kinds of traffic the controller produces into a control plane and a data plane.

\noindent \textbf{Control plane:}
The control plane carries the RESERVE/\allowbreak{}RELEASE broadcasts that synchronize the path reservation table $\mathcal{R}$ across hosts. These messages are tiny and latency-sensitive: the broadcast latency directly bounds the race window between two hosts contending for the same path. We therefore deploy the control plane on a low-bandwidth network or an existing management channel, so reservation latency is bounded by propagation rather than by queueing behind elephant flows on the data fabric. This is a deployment choice, not an algorithmic requirement: \system~still functions when control messages share the data fabric, but the race window grows under load.

\noindent \textbf{Data plane:}
The data plane carries the KV cache flows themselves over the regular RDMA fabric. On each flow arrival the controller determines $\delta_i^{*}$ and hands it to the RDMA stack, which creates the queue pair using $\delta_i^{*}$ and transmits the KV cache; switches apply standard ECMP hashing over the 5-tuple as in vanilla deployments, so no switch modification is required.

\noindent \textbf{Handling race condition:}
Because $\mathcal{R}$ is replicated and updated asynchronously, two hosts may reserve the same path within a broadcast delay window. The conflict is resolved by a deterministic timestamp-based rule: each RESERVE broadcast carries the key $(t_i,\,\mathrm{id})$, the smaller key wins, and the loser re-hashes to another path (Algorithm~\ref{alg:ark}, \textsc{OnRemoteUpdate}). The $(t_i,\,\mathrm{id})$ ordering relies on the clock-synchronization assumption stated in \S\ref{sec:approach}.
Existing approaches, such as sub-$\mu$s precision via PTP~\cite{ieee1588}, Sundial~\cite{li2020sundial}, and Huygens~\cite{geng2018huygens}, more than meet this requirement.

%% file: evaluation.tex
\section{Experiments and Results}
\label{sec:evaluation}

We evaluate \system~by integrating an LLM inference serving simulator with a packet-level network simulator.

Our evaluation answers the following questions:
\begin{itemize}[leftmargin=*]
    \item \textbf{Q1:} Does \system\ reduce per-flow flow completion time (FCT) compared to vanilla ECMP, and how does the gap scale with load? (\S\ref{subsec:multi-flow})
    \item \textbf{Q2:} Does \system\ improve fabric utilization by distributing flows more evenly across spines? (\S\ref{subsec:multi-flow})
    \item \textbf{Q3:} How do network-level FCT improvements translate to user-visible end-to-end serving latency (TTFT)? (\S\ref{subsec:4mode})
    \item \textbf{Q4:} What is \system's runtime overhead, and is it negligible compared to KV transmission time? (\S\ref{subsec:computation_time})
\end{itemize}

\subsection{Experimental Setup}
\label{subsec:experimental-setup}

\noindent \textbf{Simulation platform:}
SimAI~\cite{wang2025simai} models compute through Vidur~\cite{agrawal2024vidur}, which predicts prefill and decode latency from RandomForest regressors trained on per-layer LLaMA-3-8B profiles measured on an NVIDIA H100 GPU (varying token count, batch size, and prefill chunk size).
KV cache transfers are simulated at packet level on the ECMP-based NS-3 backend from AstraSim~\cite{rashidi2020astrasim}, on top of which we implement the \system~path reservation controller as a host-side application.

\noindent \textbf{Topology:}
We use a two-tier spine-leaf Clos topology~\cite{fattree}, matching the architecture published by Alibaba HPN~\cite{qian2024hpn}. 
The specific instance we evaluate has 16 GPUs across two 8-GPU servers, two leaf switches, and four spines, with each GPU-leaf and leaf-spine link running at 100~Gbps. This yields 800~Gbps aggregate ingress per leaf against 400~Gbps egress to the spines, creating a 2:1 oversubscription ratio that amplifies the impact of ECMP collisions under elephant-flow workloads.

\noindent \textbf{Workload:}
We model disaggregated LLaMA-3-8B inference (chosen for its public H100 latency profile) with Server~1 (GPU 1-8) as prefill nodes and Server~2 (GPU 9-16) as decode nodes; each flow is assigned a random prefill-decode pair. Prompt token counts are drawn from a Zipf distribution with $\alpha = 0.7$, capturing the heavy-tailed prompt-size pattern of real LLM serving traces and emphasizing the elephant-flow regime \system\ targets. The mean of $\sim$13{,}300 tokens yields KV cache flows from 125~MB to 6.7~GB (mean 1.6~GB) at 128~KB per token~\cite{cheng2025lmcache}. Requests arrive as a Poisson process with varying queries per second (QPS). Each experiment contains 1000 requests.

\subsection{Multi-Flow Evaluation with Realistic Workload}
\label{subsec:multi-flow}

To evaluate the effectiveness of \system, we first conduct experiments individually on the NS-3 network backend, with either \system or ECMP configured. We replay the LLM inference request trace and inject the request to NS-3 as flows, where the flow size is equivalent to the KV cache that the request generates. 

\noindent \textbf{QPS Sensitivity Analysis:}
\begin{figure}[t]
    \centering
    \begin{subfigure}[b]{0.495\columnwidth}
        \centering
        \includegraphics[width=1\linewidth]{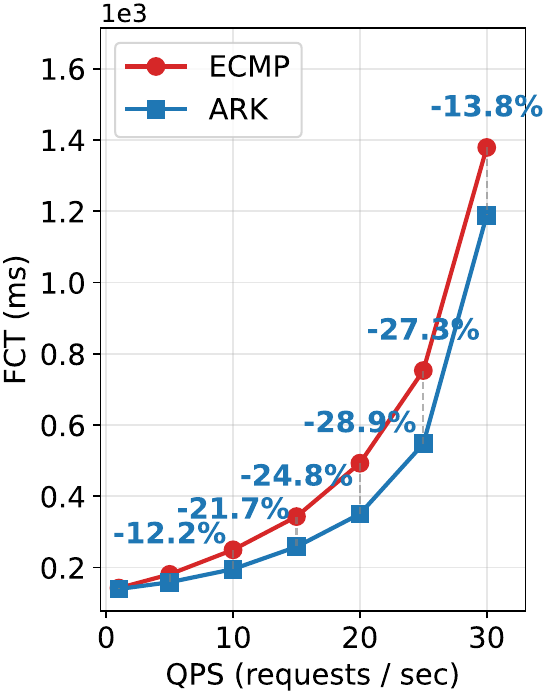}
        \caption{Mean FCT}
        \label{fig:qps_sweep_mean}
    \end{subfigure}
    \hfill
    \begin{subfigure}[b]{0.495\columnwidth}
        \centering
        \includegraphics[width=1\linewidth]{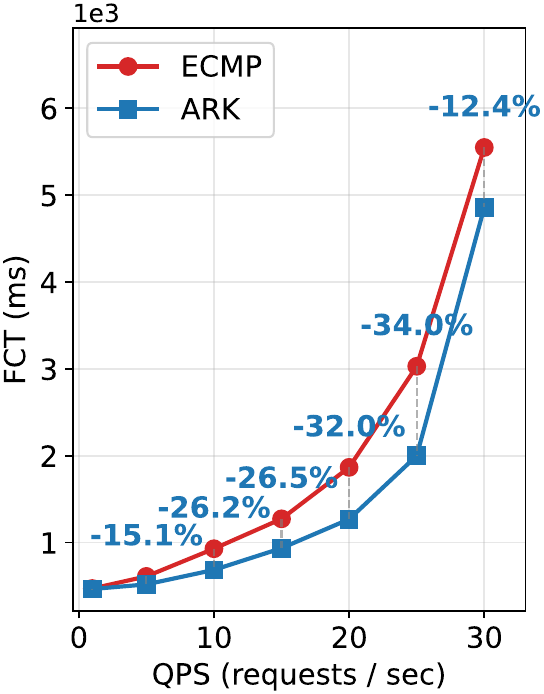}
        \caption{P95 FCT}
        \label{fig:qps_sweep_p95}
    \end{subfigure}
    \caption{Mean and P95 flow completion time (FCT) as a function of request rate (QPS) under Poisson arrival. \system consistently reduces both mean and tail FCT, with the largest improvement at QPS~=~25 (mean FCT~$-$~27.3\%, P95~$-$~34.0\%).}
    \Description{Two line plots overlaid on the same axes. The y-axis shows FCT in milliseconds (mean and P95), and the x-axis shows request rate from 1 to 30 QPS. Both metrics rise with QPS for ECMP and \system; the \system curves consistently sit below the ECMP curves, with the gap widening as QPS increases.}
    \label{fig:qps_sweep}
\end{figure}
We sweep the request rate from QPS~=~1 to 30 under Poisson arrival (Fig.~\ref{fig:qps_sweep}).
At QPS~=~1, flows rarely overlap and ECMP collisions are unlikely, so \system provides only marginal FCT reduction ($-$1.9\% mean, $-$1.2\% P95).
As QPS increases, \system delivers progressively larger improvements by distributing elephant flows across spines, peaking at QPS $=$ 25 with 27.3\% mean and 34.0\% P95 reduction. At QPS~=~30 the leaf--spine fabric approaches saturation, so the benefit narrows to 13.8\% mean / 12.4\% P95.

\noindent \textbf{Spine Utilization:}
\begin{figure}[t]
    \centering
    \includegraphics[width=\columnwidth]{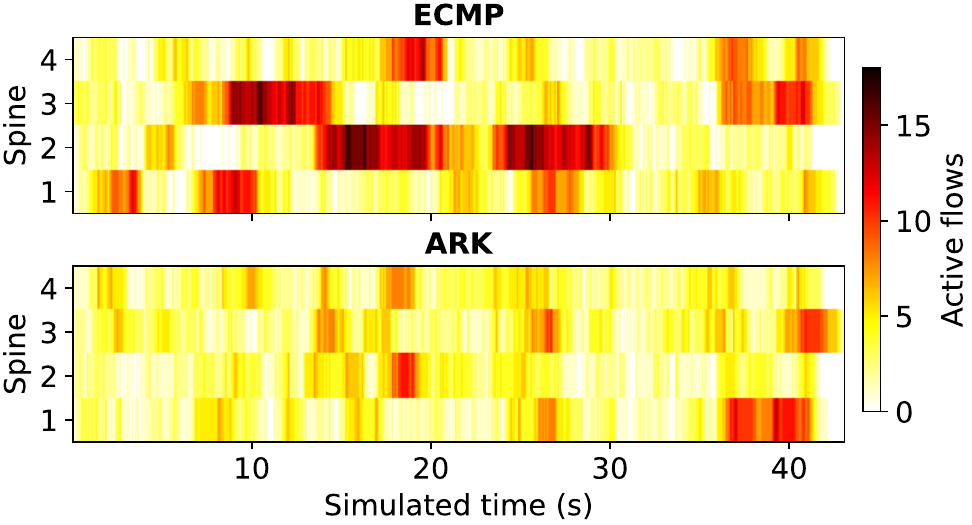}
    \caption{Per-spine active-flow heatmap at QPS~=~25 (top: ECMP, bottom: \system). ECMP drives persistent hotspots of up to 18 concurrent flows onto Spines~2 and~3 while Spines~1 and~4 stay lightly loaded; \system keeps peaks below ${\sim}$8 across all four spines.}
    \label{fig:spine_heatmap}
\end{figure}
Fig.~\ref{fig:spine_heatmap} visualizes the number of concurrent flows on each spine switch over time at QPS~=~25. With ECMP, hash collisions concentrate load on Spines~2 and~3 (peaking at 18 concurrent flows) while Spines~1 and~4 remain intermittently idle. This imbalance directly drives the FCT degradation observed in Fig.~\ref{fig:qps_sweep}, as flows routed onto a congested 100~Gbps spine experience queueing delays proportional to the number of competing flows sharing that link. \system, in contrast, spreads load uniformly across all four spines, leading to hotspot reduction and 34.0\% P95 FCT improvement.

\subsection{End-to-End Serving Evaluation}
\label{subsec:4mode}

\begin{figure}[t]
    \centering
    \begin{subfigure}[b]{0.495\columnwidth}
        \centering
        \includegraphics[width=1\linewidth]{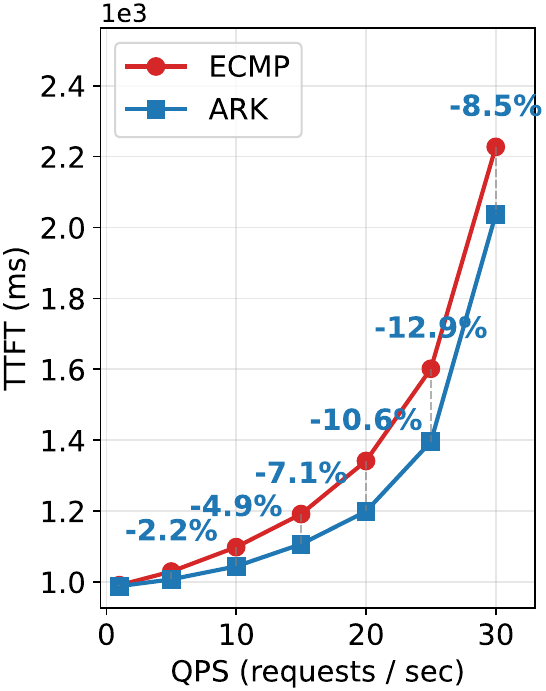}
        \caption{Mean TTFT}
        \label{fig:ttft_mean}
    \end{subfigure}
    \hfill
    \begin{subfigure}[b]{0.495\columnwidth}
        \centering
        \includegraphics[width=1\linewidth]{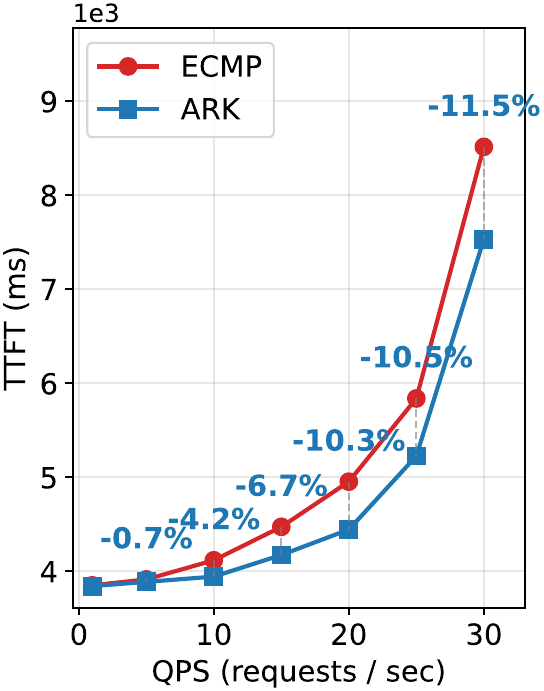}
        \caption{P95 TTFT}
        \label{fig:ttft_p95}
    \end{subfigure}
    \caption{TTFT across the same QPS sweep as Fig.~\ref{fig:qps_sweep}.}
    \Description{Two line plots sharing a common QPS axis from 1 to 30. The left panel shows mean TTFT and the right panel shows P95 TTFT, both rising with load. The ARK curve sits below ECMP at every QPS, with the gap widening from a few percent at low load to roughly twelve percent at QPS twenty-five for the mean and twelve percent for P95 at QPS thirty.}
    \label{fig:ttft_sweep}
\end{figure}

We translate the network-layer FCT improvements into user-visible serving latency. For each request,
\[
\text{TTFT} \;=\; T_{\text{prefill}}(L) \;+\; T_{\text{KV}} \;+\; T_{\text{decode-1}}(L),
\]
where $L$ is the prompt token count, $T_{\text{KV}}$ is the per-flow NS-3 FCT, and $T_{\text{prefill}}$, $T_{\text{decode-1}}$ are predicted from the Vidur Llama-3-8B with H100 profiling table.

Figure~\ref{fig:ttft_sweep} extends the QPS sweep of Fig.~\ref{fig:qps_sweep} from network FCT to user-visible TTFT. The advantage of \system widens steadily with load: a marginal 2\% reduction in mean TTFT at QPS~$=$~5 grows to roughly 13\% on the mean (QPS~$=$~25) and 12\% on the P95 tail (QPS~$=$~30). Prefill ($\sim$840~ms) and the first decode step ($\sim$10~ms) contribute a fixed offset to every request; only $T_{\text{KV}}$ scales with QPS, so as load rises the relative TTFT gain converges toward the FCT gain. TPOT, dominated by GPU-local per-token decode, is unaffected by the routing scheme.

\subsection{Computation Time} \label{subsec:computation_time}

\begin{table}[t]
\centering
\caption{\system's controller microbenchmarks on an Intel~i9 CPU (single thread, single core), and a reference KV transmission time.}
\label{tab:computation_time}
\begin{tabular}{lr}
\hline
\textbf{Phase / Operation} & \textbf{Time} \\
\hline
\multicolumn{2}{l}{\emph{Microbenchmarks (per call)}} \\
\quad Hash computation                                & 4.10~$\mu$s \\
\quad Reservation-table lookup                        & 0.87~$\mu$s \\
\hline
\multicolumn{2}{l}{\emph{Boot phase (one-time per peer)}} \\
\quad \textsc{Initialize} ($\delta_{\max} = 2^{16}$)  & $\sim$326~ms \\
\hline
\multicolumn{2}{l}{\emph{Runtime per flow (worst case)}} \\
\quad \textsc{OnFlowArrival} at $K = 4$ paths         & $\leq$3.48~$\mu$s \\
\quad \textsc{OnFlowArrival} at $K = 1000$ paths      & $\leq$870~$\mu$s \\
\hline
\multicolumn{2}{l}{\emph{Reference}} \\
\quad 1.6~GB KV cache transfer at 100~Gbps             & $\sim$137~ms \\
\quad $K{=}4$ reservation / KV transfer ratio          & $\approx$0.0025\% \\
\hline
\end{tabular}
\end{table}

We measure \system's path reservation controller on an Intel i9 CPU (single thread, single core); Table~\ref{tab:computation_time} summarizes the results. Hash computation takes 4.10~$\mu s$ per call and a path-table lookup 0.87~$\mu s$. \Call{Initialize}{} populates Eq.~\eqref{eq:precompute} with $\delta_{\max} = 2^{16}$ src-port offsets per peer at $\sim$326~ms per peer, paid once at boot. At runtime, \Call{OnFlowArrival}{} probes at most $K$ paths in $\mathcal{R}$; the $\mathcal{M}$ access is a static array dereference. The per-flow cost is therefore at most $4 \times 0.87 = 3.48~\mu s$ at $K = 4$ and scales linearly to $\sim$870~$\mu s$ at $K = 1000$, well under 1\% of the $\sim$137~ms needed to transmit a 1.6~GB KV cache flow at 100~Gbps. The controller overhead is therefore negligible compared to KV cache transmission and supports real-time KV cache flow orchestration.

%% file: conclusion.tex
\section{Conclusion}
Prefill--decode disaggregation in modern LLM serving has elevated KV cache transfers to a scale that traditional data-center fabrics were never designed to absorb. We presented \system, a distributed collision-avoidance framework that pairs (i)~a path-reservation controller steering KV cache flows onto non-overlapping spines to reduce per-flow FCT and improve spine utilization, with (ii)~a deployment that separates lightweight reservation broadcasts from KV cache transfers, enabling deployment on commodity switches. Experiments show \system~reduces mean FCT by up to 27.3\% and P95 FCT by 34.0\%, translating to up to 12.9\% mean TTFT reduction in end-to-end serving.

%% file: ethics.tex
\section*{Ethics Statement}

This work does not raise any ethical concerns. It relies entirely on
packet-level network simulation with synthetically generated workloads, and
involves no human subjects, user data, or experiments on production systems.